\documentclass[fleqn,10pt,twocolumn]{wlscirep}

\usepackage{hyperref}
\hypersetup{colorlinks=true, urlcolor= blue, citecolor=blue, linkcolor= blue, 
	bookmarks=true, bookmarksopen=false}

\usepackage{textcomp}
\usepackage{multirow}
\usepackage{soul}

\usepackage{amsmath,amssymb}
\usepackage{mathptmx}

\title{Prediction of 
	\textcolor{black}{hydrogenated} group 
	IV-V hexagonal binary monolayers}

\author[1,+]{Mohammad Ali Mohebpour}
\author[1,+]{Shobair Mohammadi Mozvashi}
\author[2]{Sahar Izadi Vishkayi}
\author[1,*]{Meysam Bagheri Tagani}
\affil[1]{Computational Nanophysics Laboratory 
	(CNL), Department of physics, University of Guilan, P. O. Box 
	41335-1914, Rasht, Iran.}
\affil[2]{School of Physics, Institute for Research in 
	Fundamental Sciences (IPM), P. O. Box 19395-5531, Tehran, Iran.}

\affil[*]{m{\_}bagheri@guilan.ac.ir}

\affil[+]{these authors contributed equally to this work}

\begin{abstract}
Group IV and V 
monolayers are very crucial 2D materials for their high carrier
mobilities, tunable band gaps, and optical linear dichroism. Very recently, 
a
novel group IV-V binary compound, Sn$_2$Bi, has been 
synthesized on silicon
substrate, and has shown very interesting electronic properties. Further 
investigations have revealed that the  monolayer would
be stable in freestanding form by hydrogenation. Inspired by this, by means 
of 
first-principles
calculations, we systematically predict and investigate eight
counterparts of
Sn$_2$Bi, namely Si$_2$P, Si$_2$As, Si$_2$Sb, Si$_2$Bi, Ge$_2$P, Ge$_2$As,
Ge$_2$Sb, and Ge$_2$Bi.
The cohesive energies, phonon dispersions, and AIMD calculations show that,
similar to Sn$_2$Bi, all of these freestanding
monolayers are stable in hydrogenated form. These
hydrogenated monolayers are semiconductors with wide band gaps, which are
favorable for opto-electronic purposes. The Si$_2$YH$_2$ and
Ge$_2$YH$_2$ structures possess indirect and direct band gaps, respectively.
They represent very interesting optical characteristics,
such as good absorption in the visible region and linear dichroism, which 
are
crucial for solar cell and beam-splitting devices, respectively. Finally,
the Si$_2$SbH$_2$ and Si$_2$BiH$_2$ monolayers have suitable band 
gaps and 
band edge
positions for photocatalytic water splitting. Summarily, our investigations
offer very interesting and promising properties for this family of
binary compounds. We hope that our predictions open
ways to new experimental studies and fabrication of suitable 2D materials 
for
next generation opto-electronic and photocatalytic 
devices.
\end{abstract}
\begin{document}

\flushbottom
\maketitle
%
%
\thispagestyle{empty}

\section{Introduction}
The high tower of the contemporary technology is built by blocks of silicon
and germanium.
Since the successful synthesize of the monolayer carbon (graphene)
\cite{novoselov04} and
discovery of its remarkable characteristics, such as high carrier mobility
\cite{hwang08},
strong mechanical parameters \cite{frank07}, and optical transparency
\cite{nair08}, a great inquiry
for other elemental monolayers is in the agenda of many scientists around 
the
world. The monolayers of carbon’s neighbors in group-IV, silicon and
germanium (silicene and germanene) are among the most important predicted 
and
synthesized monolayers beyond graphene \cite{lalmi10, li14}.

Unlike graphene,
which is
completely flat with an sp$^2$ bonding characteristics, the larger 
interatomic
distance in silicene and germanene weakens the $\pi - \pi$ overlaps, which
leads to buckled structures with sp$^2$ $-$ sp$^3$ hybrid orbitals. Despite
their buckled geometry, silicene and germanene share most of the important 
electronic features of graphene, such as Dirac cone, high Fermi velocity 
and carrier mobility \cite{ezawa15, acun15}, with some advantages including 
better
tunability of the band gaps \cite{drummond12}, stronger spin-orbit coupling
\cite{liu2011}, and easier valley polarization \cite{Tabret13}, which are 
very important for electrics, spintronics, and valleytronics.

On the other hand, monolayers of group-V elements, known as pnictogens,
including phosphorene, arsenene, antimonene, and bismuthene, recently have
gained much attention for their topological aspects, as well as inherent, 
wide,
and tunable band gaps \cite{zhang16, kamal15, Wang15, akturk16}. Generally,
several allotropes are considered for
these monolayers, including $\alpha$ (puckered or washboard) and $\beta$ 
(buckled honeycomb or
graphene-like), as the most important and stable phases. For
arsenene, antimonene, and bismuthene, the $\beta$-phase, and for 
phosphorene, the $\alpha$-phase is
more stable in aspects of energetics and phonon dispersions 
\cite{zhang16,Pumera}. 
The
$\alpha$-phase
phosphorene
and
arsenene
possess direct band gaps, while their $\beta$ counterparts have indirect 
ones.
On the
other hand, antimonene and bismuthene respectively have indirect and direct
band gaps in both phases. These band gaps are within a wide range of 0.36 
(for $\alpha$-bismuthene) to 2.62 eV (for $\beta$-phosphorene) 
\cite{akturk16, kamal15, Wang15}.
Moreover,
phosphorene, arsenene, and bismuthene possess carrier mobilities as high as
several thousand  \mbox {cm$^2$ V$^{-1}$  s$^{-1}$} \cite{zhang16}. These
exciting
properties
makes
group-V
monolayers very favorable candidates for optoelectronics, 
and photocatalytic devices.

Because of high ratio between the surface and thickness of 2D structures,
effects of chemical functionalization play an important role in tuning their
properties. Hence, in addition to pure elemental
monolayers, 2D materials with functionalized structures gained attention for
expanding the scope of realized physical aspects and enhancing
potential applications. These efforts include designing and applying various
types of heterostructures \cite{mozvashi19}, 
defections\cite{bafekry19}, 
vacancies \cite{hadipour19}, adsorptions 
\cite{Tagani18}, and
compounds \cite{Shafique, mahmood16}. Among these, binary compounds have the 
advantage of
relatively easier fine control of the growth dynamics and more feasible
fabrication. They could represent unusual atomic configuration and chemical
stoichiometry \cite{gou18} which leads to extraordinary physical
properties for future applications and opening ways to new researches.

As an example of group IV-V binary compound, Barreteau et al have succeed to
synthesize the bulk single crystals of layered SiP, SiAs, GeP, and GeAs by
melt-growth method. They showed that these layered materials all exhibit
semiconducting behavior, and suggest that they can be further exfoliated 
into 2D structures \cite{barreteau16}. Moreover, a number of recent 
theoretical works were performed on group \mbox{IV-V} 2D binary compounds and 
reported 
interesting results in thermoelectricity for SiX (X=N, P, As, Sb, and 
Bi)\cite{somaiya2020}, visible-light photohydrolytic 
catalysts for SiP \cite{ma2018sip}, strain-tunable electron mobility for XY (X 
= C, Si, and Ge, and Y = N, P, and As) \cite{zhang2018strain}, 
and ORR applications in novel fuel cells for metal (Ni, Pd, Pt,
and Ru) complexes in graphene basal planes \cite{zhang2020mechanism}. 

Very recently, Gou et al. have synthesized a unique hexagonal 2D binary
compound, Sn$_2$Bi,
on a silicon (111) substrate which exhibits strong spin-orbit coupling and 
high
electron-hole asymmetry \cite{gou18}. In the band structure of this
semiconducting monolayer, electron flat bands and free hole bands are seen
which are indicatives of nearly free and strongly localized charge carriers.
Moreover, this monolayer is very stable because all the Si, Bi, and Sn atoms
satisfy the octet rule. These features make Sn$_2$Bi a good candidate for
nano-electronics and may result in nontrivial properties like ferromagnetism
\cite{mielke99} and superconductivity \cite{Shen}. Furthermore, the 
synthesis
of other group-IV-V X$_2$Y counterparts of Sn$_2$Bi was proposed by Gou et al 
\cite{gou18}.

Generally, experimental synthesis of yet unknown
systems can be guided by predictive theoretical first-principles
calculations which distinguish stable and unstable structures correctly. In 
other words, theoretical predictions play an 
important
role in progress of materials science and technology, by means of
justifying the cost and effort of potential experiments.
Many advances in materials science have been conducted and inspired by 
earlier
theoretical investigations. Most of the presently well-known synthesized 2D
materials,
such as borophene \cite{boustani97}, stanene,
germanene, silicene \cite{ezawa15}, arsenene \cite{kamal15}, antimonene 
\cite{Wang15}, bismuthene
\cite{akturk16}, 
etc. were firstly predicted by theoretical studies which brought sufficient
motivations for experimental work.

Herein, inspired by the successful deposition of Sn$_2$Bi monolayer, as 
well as
the importance of group IV and V monolayers, we predicted a new family of
binary
compound monolayers with a hexagonal structure and an empirical formula of
X$_2$Y,
where X and Y are respectively chosen from group-IV (Si and Ge) and V (P, 
As,
Sb, and Bi), namely Si$_2$P, Si$_2$As, Si$_2$Sb, Si$_2$Bi, Ge$_2$P, 
Ge$_2$As,
Ge$_2$Sb, and Ge$_2$Bi. We firstly stabilize the mentioned monolayers by
hydrogenation, and further
check their stability by cohesive energy, molecular dynamics, and phonon
dispersion analysis, and interpret their phonon modes and thermodynamical
properties. Furthermore, we analyze their electronic and optical properties 
and discuss their potential strengths. Eventually, we consider these
semiconductors for photocatalytic purposes and check their potential
applications in water-splitting.

Our results suggest that these monolayers 
are
strongly applicable in a very vast areas such as valleytronics,
opto-electronics, beam-splitters,
optical detectors, and
water-splitters. Moreover, the structural similarity with the synthesized
Sn$_2$Bi monolayer, promises the possibility of their deposition on proper 
substrates and brings hopes for advances in technological devices.

\section{Computational Details}

The first-principles calculations were performed based on the density
functional theory (DFT), as implemented in the Quantum Espresso package
\cite{giannozzi09}.
During the entire calculations, the norm-conserving (NC) pseudo-potentials 
with
a \textcolor{black}{plane} wave basis set were employed to describe the
electron
wave functions. The generalized gradient approximation (GGA) was used with 
the
formulation of Perdew-Burke-Ernzerhof (PBE) to describe the 
exchange-correlation potential \cite{perdew96}. Because the GGA usually
underestimates the band gaps, the HSE06 hybrid functional was also used to
obtain more
accurate band gaps. \textcolor{black}{The energy cut-off for wave function 
	and 
	charge density was 
	set to 50 and 300 Ry, respectively. The 
	Monkhorst-Pack scheme was used to sample the Brillouin 
	zone with a 13$\times$13$\times$1 and 21$\times$21$\times$1 
	\mbox{k-points}  
	for 
	geometric optimization and electronic calculations, respectively.  However, 
	for 
	the HSE 
	calculations, the \mbox{k-points} 
	was set to be 5$\times$5$\times$1.} A vacuum space of 20 \AA\ was chosen 
	along 
the z-direction to
prevent spurious interactions between layers in the periodic boundary
condition. All the monolayers were fully relaxed with a force and stress
tolerance of 10$^{-3}$ eV/\AA\ and 10$^{-4}$~GPa, respectively. To calculate
the phonon dispersion, the finite displacement method was adopted, in which
a
3$\times$3$\times$1 supercell with a 5$\times$5$\times$1 k-point sampling 
was
built.

\textcolor{black}{To investigate the optical properties, the 
	frequency-dependent 
	dielectric function was calculated within the independent particle 
	approximation (IPA) which describes single-particle excitations, as 
	implemented 
	in the epsilon code inbuilt in the QE package. The calculation was 
	performed by 
	means of self-consistent ground-state eigenvalues and eigenfunctions.}

To determine the structural stability of the monolayers, their cohesive
energies
($E_c$) were calculated using the equation below:

\begin{equation}\label{coh}
E_c=\frac{E_{sheet}-\sum_{i} n_i E_{atom-i}}{N}
\end{equation}

\noindent where $E_{sheet}$ and $E_{atom-i}$ stand for total 
energy of the
sheet
and the isolated atom-i with considerations of the spin polarization,
respectively. $N$ and $n_i$ are the numbers of total atoms and atom-i in
the unit cell, respectively.

To check the thermal stability, the ab-initio molecular
dynamics (AIMD) simulations were performed using NVT canonical ensemble at 
room
temperature (300 K). The initial model was constructed by a 
3$\times$3$\times$1
supercell for
minimizing the constraint caused by periodicity. Here, the total simulation
time was set to be 4.0 ps with time steps of 2.0 fs.
\section{Results and Discussion}
\subsection{Structural Stability and Phonon Calculations}

\begin{table*}[t]
	\centering
	\caption{Structural parameters of the X$_2$YH$_2$ binary
		compound monolayers, including lattice constants ($a$), bond lengths
		($R_{X\!X}$ and $R_{X\!Y}$), buckling heights ($\Delta$),
		cohesive energies
		($E_c$), band gaps ($E_g$), Debye temperatures ($\theta_D$), and 
		constant
		volume heat capacity in room temperature ($C_V^{300 K}$).}
	\label{configtab}
	
	\begin{tabular}{lcccccccc}
		\hline
		& $a$ (\AA) & {$R_{X\!X}$ (\AA)}&$R_{Y\!Y}$ (\AA) &
		$\Delta$
		(\AA) &
		$E_c$
		(eV/atom) &
		$E_g:$ GGA, HSE (eV)& $\theta_D$ (K)&$C_V^{300 K}$ (J~mol$^{-1}$
		K$^{-1}$) \\
		\hline
		Si$_2$PH$_2$& 6.26 & 2.35 & 2.27 & 1.08 & -3.88 & 2.39, 3.19 (ind) &
		143.7&15.45\\
		Si$_2$AsH$_2$& 6.44 & 2.35 & 2.39 & 1.19 & -3.73 & 2.33, 3.04 (ind) &
		120.3&16.09\\
		Si$_2$SbH$_2$& 6.79 & 2.35 & 2.60 & 1.30 & -3.58 & 2.04, 2.61 (ind) &
		97.9&16.57\\
		Si$_2$BiH$_2$&6.94 & 2.35 & 2.69 & 1.35 & -3.50 & 1.92, 2.43 (ind) & 
		70.8&
		16.93\\
		Ge$_2$PH$_2$&6.52 & 2.46 & 2.36 & 1.15 & 3.34 & 2.21, 2.88 (dir) &
		94.7&17.18\\
		Ge$_2$AsH$_2$& 6.69 & 2.47 & 2.47 & 1.23 & -3.24 & 1.80, 2.41 (dir) &
		83.5&17.71\\
		Ge$_2$SbH$_2$&7.03 & 2.47 & 2.67 & 1.33 & -3.13 & 1.57, 2.07 (dir) &
		69.3&18.03\\
		Ge$_2$BiH$_2$&7.18 & 2.48 & 2.75 & 1.38 & -3.08 & 1.17, 1.57 (dir) &
		54.9&18.22\\
		\hline
	\end{tabular}
\end{table*}

\begin{figure*}[h]
	\centering
	\includegraphics[width=\textwidth]{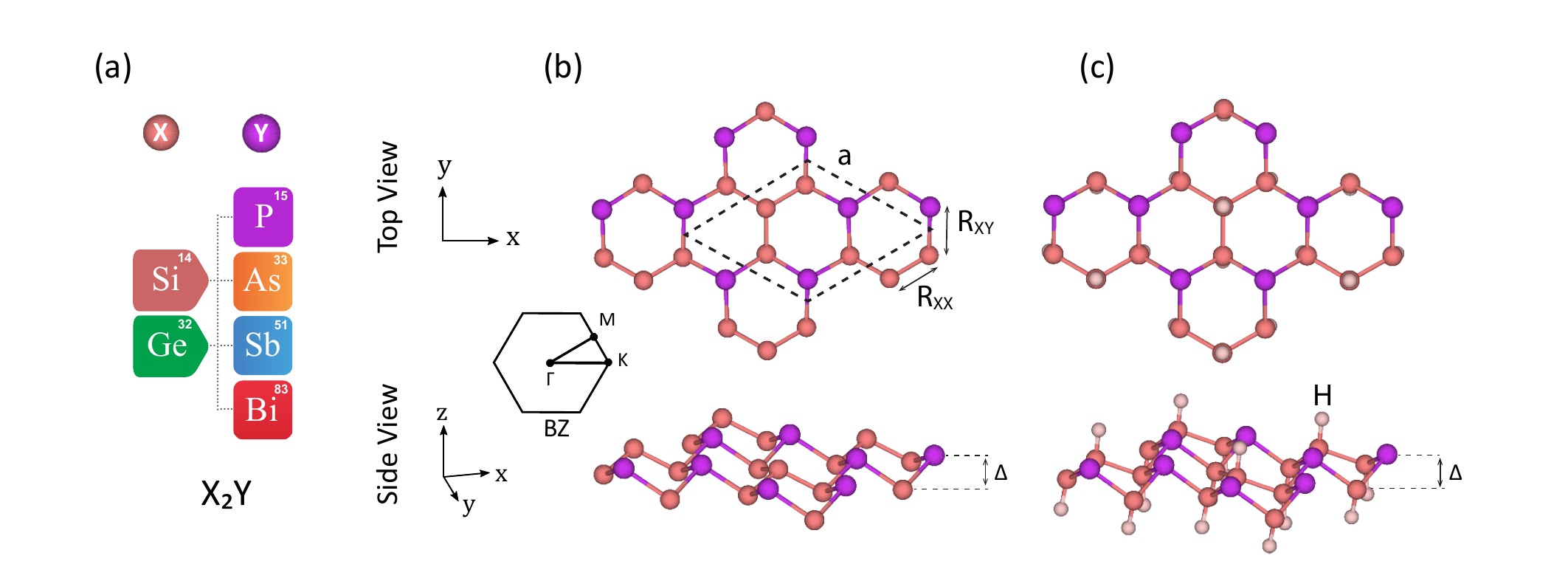}
	\caption{(Color online) Structural configurations of the predicted 
		binary
		compound monolayers. \textbf{(a)} Table of included elements. 
		\textbf{(b)}
		Top
		and \textbf{(c)} side view of the pure and hydrogenated monolayers. 
		The
		unit cell and the corresponding Brillouin zone have also been 
		presented.}
	\label{config}
\end{figure*}

\begin{figure}[h!]
	\centering
	\includegraphics[width=0.4\textwidth]{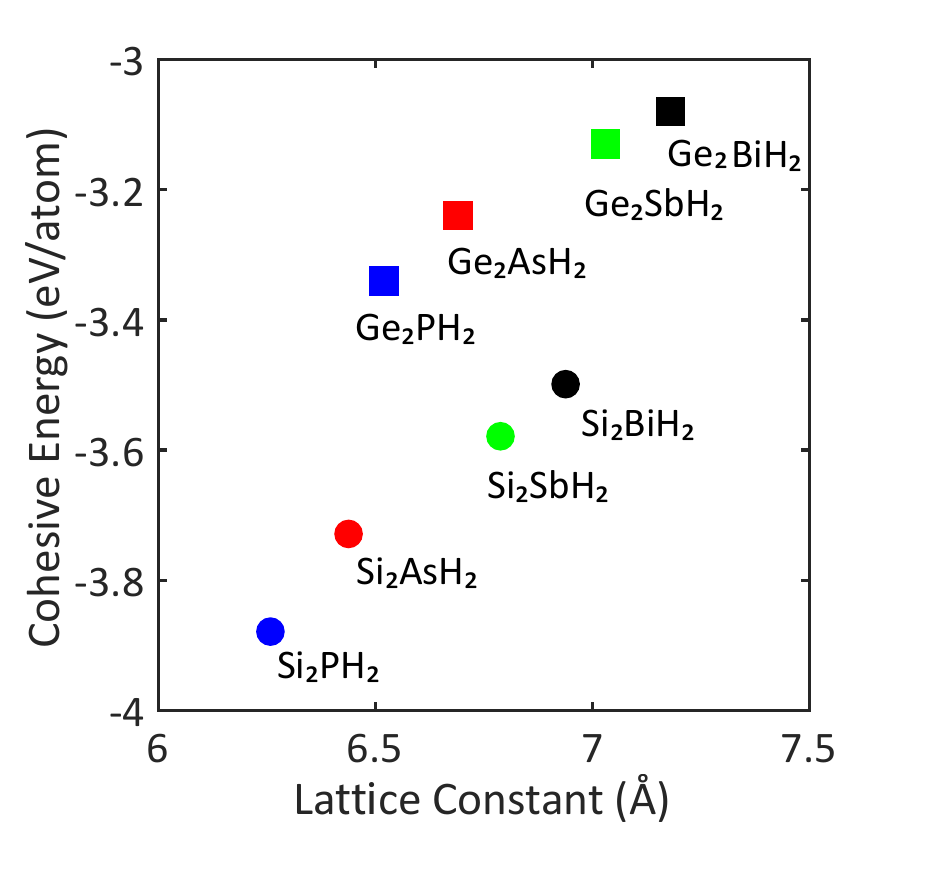}
	\caption{Variation of cohesive energy with lattice constant of the
		X$_2$YH$_2$ binary compound monolayers.}
	\label{cohlatt}
\end{figure}

\begin{figure}[h!]
	\centering
	\includegraphics[width=0.5\textwidth]{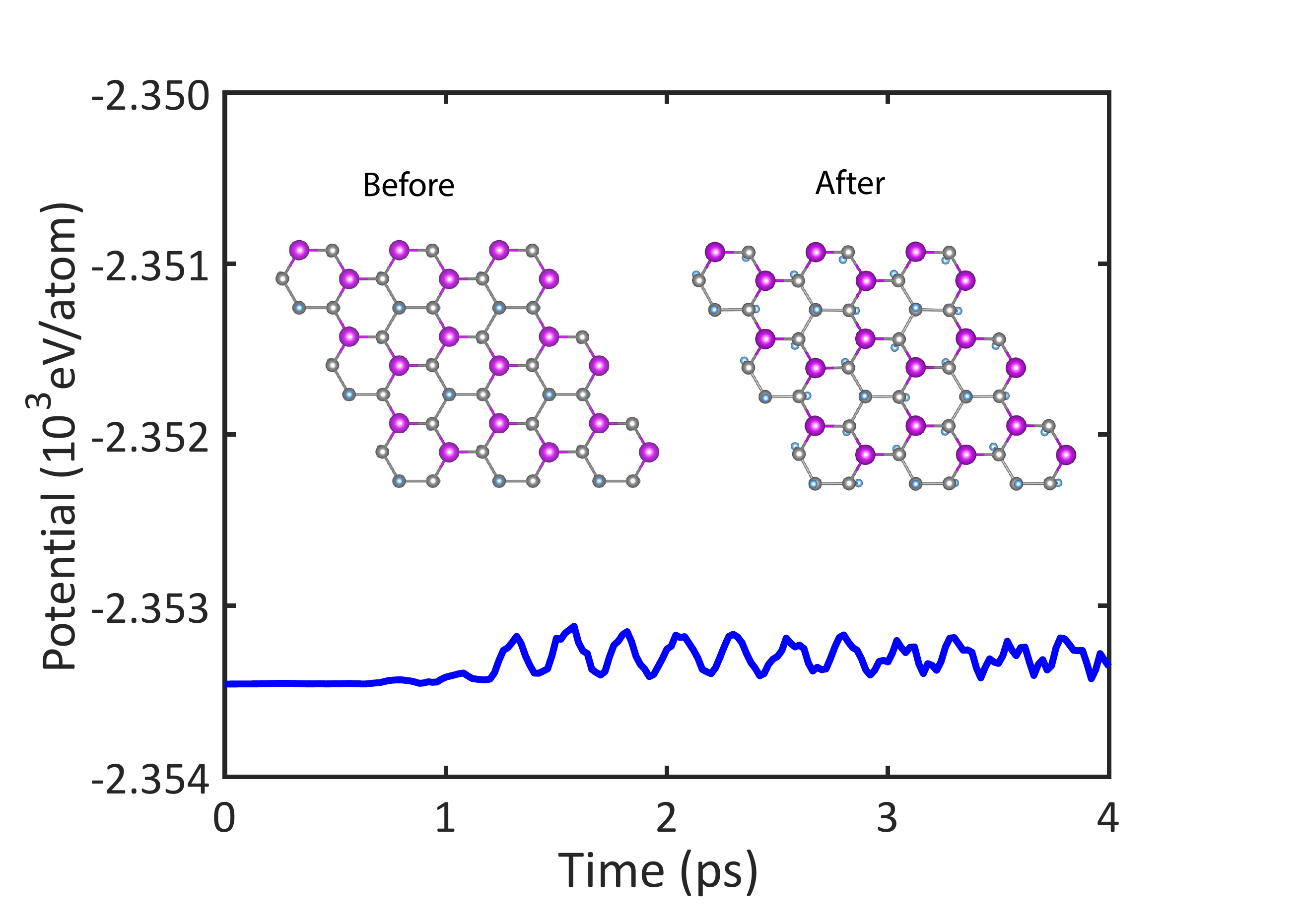}
	\caption{Potential energy fluctuations of the Ge$_2$BiH$_2$ 
		during
		the
		AIMD simulations at 300 K. The final geometric structure at the end 
		of 4 ps
		has also been shown.}
	\label{MD}
\end{figure}

\begin{figure*}[t]
	\centering
	\includegraphics[width=\textwidth]{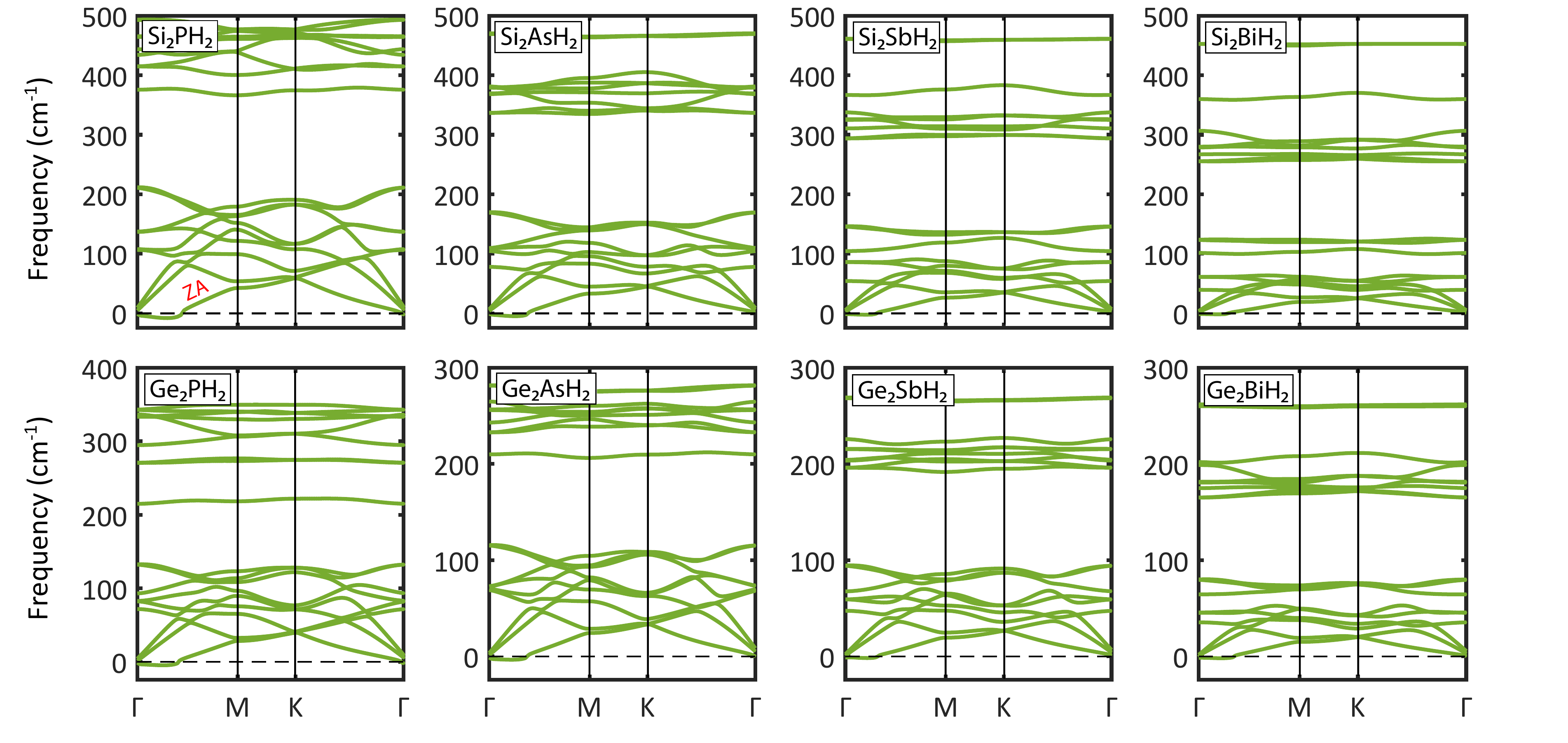}
	\caption{Phonon dispersion spectra of the X$_2$YH$_2$ binary
		compound monolayers. As can be seen, there are 18 phonon branches
		corresponding to 6 atoms in the unit cell (excluding hydrogen 
		atoms). No
		considerable imaginary modes are seen, so all the structures are
		dynamically stable.}
	\label{phonon}
\end{figure*}

\begin{figure}[h]
	\centering
	\includegraphics[width=0.5 \textwidth]{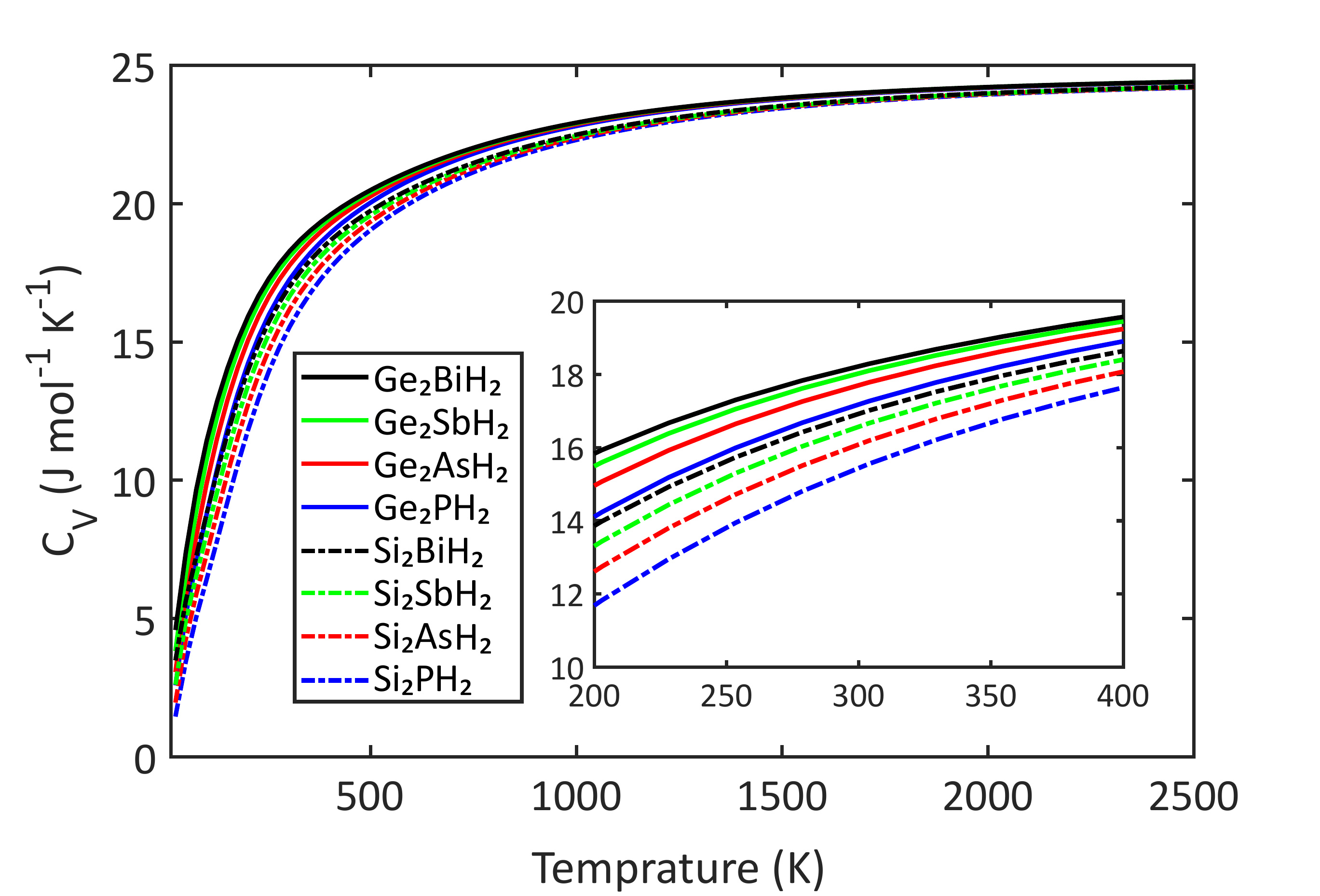}
	\caption{(Color online) Constant volume heat capacity ($C_V$) of the
		predicted binary compounds as a function of temperature, for one 
		mole, and
		divided by the number of atoms in the unit cell (10). The $C_V$ is
		converged
		to $\sim$ 24 J mol$^{-1}$ K$^{-1}$ in high-temperature limit, which 
		is in
		agreement with the Debye model. Also, the heavier monolayers have 
		greater
		$C_V$ at room temperature, which is consistent with similar 
		studies.}
	\label{CV}
\end{figure}

The graphene-like structure was used to construct eight new binary 
monolayers,
with a threefold-coordinated X (Si and Ge) and Y (P, As, Sb, and Bi) atoms 
in a
hexagonal unit cell containing
six atoms, as shown in Fig.~\ref{config}. Through the structural 
optimization
with the GGA-PBE exchange-correlation, the relaxed lattice constants and
bond lengths were calculated in the range of \mbox{6.33 to 7.23 \AA\ } and
2.26 to 2.75 \AA,
respectively. All the relaxed monolayers have buckled structures with 
buckling
heights in the range of 0.86 to 1.24 \AA\ in which the longer atomic radius
creates larger buckling heights. Moreover, all of the monolayers represent 
metallic electronic properties. The calculated structural parameters are
available in the Supporting Information, Table S1.

The X and Y atoms have ns$^2$~np$^2$ (n = 3, 4) and
ns$^2$~np$^3$ (n = 3 $-$ 6)
outer shell
electron configurations, respectively. Therefore, when they form a threefold
configuration, the octet rule only fulfills for the Y, not X atoms. Thus, 
these
pure structures are predicted to be unstable in a freestanding 
configuration.
Phonon dispersion analyses confirm that these monolayers are dynamically
unstable (Fig. S1).
The same instability has also been reported for freestanding Sn$_2$Bi
monolayer while it can be greatly stabilized by hydrogenation \cite{ding19,
	mohebpour20}.

\textcolor{black}{Hydrogenation 
	is a well-recognized technique for stabilization and tuning the physical 
	characteristics of nano-scale systems. 
	Many experimental studies have been done on performing different methods 
	of hydrogenation. For 
	example, the surface hydrogenated graphene (aka graphane) was prepared by 
	exposure 
	of graphene to a cold hydrogen plasma which led to the opening of a band 
	gap and 
	other 
	changes in its electronic properties \cite{elias2009}. 
	This 
	success was inspired by previous theoretical predictions which conducted 
	the 
	experiment well into a new graphene-based structure \cite{sofo2007, 
		pumera2013graphane}. In addition, several phases of borophene have been 
	synthesized on Ag (111), Au (111), and Cu (111) substrates, which are metal 
	and 
	unstable in freestanding form. Eventually, a new phase of hydrogenated 
	borophene was synthesized by thermal decomposition of sodium borohydride 
	(NaBH$_4$) powders, which shows an ultra-stability and semi-conducting 
	characteristics in the air environment \cite{hou2020}. This achievement was 
	also led by theoretical predictions \cite{xu2016hydrogenated}.}

\textcolor{black}{In our case, the Sn$_2$Bi monolayer was firstly synthesized 
	on 
	a 
	silicon 
	substrate \cite{gou18}. Subsequently, a computational study 
	suggested that the isolated Sn$_2$Bi is a metal which suffers from 
	instability 
	due to the dangling bonds. However, it can become a stable semiconductor by 
	use 
	of chemical functionalization, such as surface hydrogenation \cite{ding19}. 
	By 
	comparison, it is found that the results of 
	hydrogenated Sn$_2$Bi are very similar to that of Sn$_2$Bi synthesized on 
	the 
	substrate. For example, in both systems, there are electron flat bands and 
	free 
	hole bands, which provide the possibility of having strongly localized 
	electrons and free holes. Also, the bandgap predicted for hydrogenated 
	Sn$_2$Bi 
	is 0.92 eV which agrees well with bandgap of 0.8 eV for substrate-supported 
	monolayer revealed by angle-resolved photoemission spectroscopy 
	measurements 
	\cite{ding19, gou18}.
	In the following, we show that the 
	predicted X$_2$Y monolayers, similar to Sn$_2$Bi, can be stabilized and 
	become 
	semiconductors by surface hydrogenation. In 
	other words, we found an interesting analogous trend shared with Sn$_2$Bi 
	and 
	its eight counterparts suggested by us.
	Summarily, according to similarity of Sn$_2$BiH$_2$ and Sn$_2$Bi/Si(111) 
	\cite{gou18,ding19}, we predict that 
	the properties of the 
	hydrogenated X$_2$Y monolayers are also similar to 
	possible deposited monolayers on suitable insulator substrates such as ZnS 
	(111), SiC (111), and Si(111)}.

For surface hydrogenation, we investigated both single and double side 
hydrogenated
structures, where hydrogen make bonds with X (Si and Ge) atoms, so the
octet rule would be fulfilled. According to the cohesive energies, the 
double
side hydrogenated model, having the lowest ground state energy, is 
predicted to
be the most stable structure. Therefore, we denote the rest of the
investigations to
this model which is described in Fig.~\ref{config}c. In the following, we
further confirm
their structural, thermal and dynamical stability by means of cohesive 
energy,
molecular dynamics, and vibrational phonon analysis.

Table~\ref{configtab} lists the structural and electronic parameters for 
these
monolayers.
Lattice constants, bond lengths, and buckling heights are in the
range of 6.26 to \mbox{7.18 \AA}, 2.27 to 2.75 \AA, and 1.08 to 1.38 \AA,
respectively.
As can be seen, hydrogenation causes an increase in buckling
heights and a decrease in lattice constants for all the monolayers, which is
due to the strong
bonds between H and X atoms. Similar behaviors have also been reported for
hydrogenation and fluorination of
penta-graphene \cite{li16-vJcrg}, silicene \cite{goli2020}, germanene
\cite{Trivedi14}, and stanene \cite{chen16}.

We have also performed the ab-initio molecular dynamics (AIMD) simulations 
to
verify the thermal stability of the X$_2$YH$_2$ binary compounds.
Fig.~\ref{MD} exhibits the fluctuations of potential energy and
evolutions of geometric structure of the Ge$_2$BiH$_2$ monolayer 
during
the
simulations
at 300 K. As can be seen, the potential energy oscillates with an extent of
less than 0.4 eV/atom, and no obvious structural distortions are found,
indicating that the Ge$_2$BiH$_2$ is thermally stable at 300 K. The
thermal stability of the Ge$_2$BiH$_2$ guarantees stability of all 
the
predicted structures because it has the highest cohesive energy among them
(see
Fig.~\ref{cohlatt}). Indeed, this suggests that the X$_2$YH$_2$
binary compounds can be realized experimentally at room temperature.

To further confirm the stability of the hydrogenated monolayers, the phonon
dispersion spectra were calculated and displayed in Fig.~\ref{phonon}. It 
is
clear that
there is no imaginary frequency in the whole Brillouin zone, which confirms
that these freestanding monolayers are dynamically stable. The spoon-shaped
curves
near the $\Gamma$ point do not mean instability, but they are signatures of 
the
flexural acoustic modes, which are usually hard to converge in 2D sheets. 
These
soft modes are also found in other analogous systems \cite{Yu16, Zheng15}.

All
the phonon spectra have rather similar trends, which mean similar bonding.
Also, it is clear that the maxima of acoustic modes decline with going down 
in
group IV and V where Si$_2$PH$_2$ and Ge$_2$BiH$_2$ display the highest (100 
cm$^{-1}$)
and lowest (38 cm$^{-1}$) peaks. Based on these maxima, the Debye 
temperatures
are obtained by $\theta_D=h\nu_m/K\!_B$ \cite{guo17}, where $h$
and
$K\!_B$
are
the Planck and Boltzmann constants, respectively. The calculated 
temperatures
are in the range of 143 to 54 K (listed in Table~\ref{configtab}) which are
lower than graphene (2266 K), silicene (798 K), phosphorene (206 K), 
arsenene
(170 K), and comparable to antimonene (101 K), bismuthene \mbox{(50 K)}, and
stanene
(72 K) \cite{guo17, Wang16, kumar17, kocabas18}. Such low Debye temperatures
and large buckling heights, which are indicatives of low lattice thermal 
conductivity, may bring hope for these monolayers to be suitable candidates for 
thermoelectric applications.

Interestingly, the slope of the parabolic out-of-plane acoustic mode (ZA) 
near
the $\Gamma$ point (specified in \mbox{Fig.~\ref{phonon}}) decreases with
increasing
of the average atomic mass of the monolayers. This will bring a slower 
phonon
group velocity, subsequently lower lattice thermal conductivity, and
stronger anharmonicity, especially for Si$_2$BiH$_2$, Ge$_2$AsH$_2$, 
Ge$_2$SbH$_2$, and
Ge$_2$BiH$_2$. It is worth noting that the ZA mode has a high
contribution to the phonon transport \cite{nika12}. On the other hand, the
hybridization of the optical and acoustic phonon branches increases the 
phonon scattering which reveals low phonon transport. These behaviors represent
the possible potential of X$_2$YH$_2$ monolayers in thermoelectricity. 

According to Eq. (\ref{coh}), the more negative values for cohesive
energies
suggest more structural stability for the monolayers. As shown in Table
\ref{configtab},
the cohesive energies vary from -3.88~eV/atom for Si$_2$PH$_2$ to 
-3.08~eV/atom 
for 
Ge$_2$BiH$_2$
which
indicates that all of the monolayers are stable. In fact, the structures
represent more stability when the atoms are lighter. By comparison, one can
easily realize that all the predicted monolayers are more stable than the
hydrogenated Sn$_2$Bi (Sn$_2$BiH$_2$), \textcolor{black}{which was discussed 
	in 
	our previous 
	study to have a 
	cohesive energy of -2.95 eV/atom \cite{mohebpour20}.}
{Also,
	the Si$_2$SbH$_2$,
	Si$_2$BiH$_2$,
	Ge$_2$SbH$_2$, and Ge$_2$BiH$_2$ monolayers are more stable than SiSb 
	\mbox{(-3.50 
		eV/atom)},
	SiBi
	(-3.31 eV/atom), \mbox{GeSb (-3.12 eV/atom)}, and GeBi (-2.98 eV/atom) 
	binary 
	compounds, 
	respectively.} The
rest have appreciable cohesive energies comparable to SiP (-4.19~eV/atom), SiAs
(-3.85~eV/atom), GeP (-3.60 eV/atom), and GeAs (-3.36~eV/atom) 
\cite{ozdamar18}. \textcolor{black}{All the mentioned cohesive energies above 
	were 
	calculated through GGA-PBE functional.} For a better
comparison between
cohesive energies of the predicted binary compounds, please pay attention to
Fig.~\ref{cohlatt}.

Phonon dispersion is also a key to calculate thermodynamic
properties of a system. For example, the constant volume heat capacity, 
$C_V$ is defined as \cite{mogulkoc19}:
\vspace{-3pt}
\begin{equation}
C_V=\sum_{s,q}K\!_B\left( \frac{\hslash\omega_s(q)}{K\!_BT}\right)^2
\frac{exp(\hslash\omega_s(q)/K\!_BT)}
{\left(exp(\hslash\omega_s(q)/K\!_BT)-1\right) ^2}
\end{equation}

\noindent where $\hslash$ is the reduced Planck's constant, and $\omega_s(q)$ 
is
the frequency of the $s$ phonon branch at the $q$ point. According to
the
Debye model, in the high-temperature limit, i.e. $K\!_BT \gg \hslash\omega$, 
the
heat capacity simply approaches to the classical Dulong-Petit results, 
which is
$3N\!M\!K\!_B$, where $N$ is the number of atoms in the unit cell and $M$ is
the number
of unit cells in a crystal ($\approx$~24.94 J~mol$^{-1}$~K$^{-1}$   for one
mole of a mono-atomic solid) \cite{kittel96}. \mbox{Fig.~\ref{CV}}
exhibits
the
$C_V$ calculated
for
the hydrogenated binary compounds as a function of temperature (one mole,
divided by the number of atoms in the unit cell) which was calculated by use
of the phonon dispersion spectra. As it is clear, the $C_V$ is converged to
$\sim$~24~J~mol$^{-1}$~K$^{-1}$ in high-temperature limit,
which is in good agreement with the Debye model.

Moreover, the $C_V$ for Si$_2$PH$_2$, Si$_2$AsH$_2$, Si$_2$SbH$_2$, 
Si$_2$BiH$_2$, Ge$_2$PH$_2$,
Ge$_2$AsH$_2$, Ge$_2$SbH$_2$, and
Ge$_2$BiH$_2$, at room temperature (300 K) are 15.45, 16.09, 16.57, 16.93, 
17.18,
17.71, 18.03, and 18.22 J~mol$^{-1}$~K$^{-1}$, respectively (see Table
\ref{configtab}). Despite the
importance of the $C_V$ in the understanding of thermal properties, it has 
not
gained sufficient attention in 2D materials so far. To the best of our 
probe,
some examples of similar calculations are: 23.1 (TiSeS), 22.7 (TiTeS), 22.5
(TiSeTe), 17.5 (CuTe$_2$O$_5$), 11.5 (borophene) J~mol$^{-1}$~K$^{-1}$ 
\cite{mogulkoc19, pekoz18,
	lysogorskiy15}, which are comparable
with our results. It is provable that heavier materials have a greater 
$C_V$ at
room temperature, i.e. they are more resistant to temperature increase.
Therefore, one may conclude that compared with borophene, all of the 
predicted
binary compounds, and compared with CuTe$_2$O$_5$, the Ge$_2$AsH$_2$,
Ge$_2$SbH$_2$, and
Ge$_2$BiH$_2$
monolayers are better electronic devices in the aspects of not overheating.
With confirming the structural stability and discussing the thermodynamical
characteristics, now we turn our attention into the electronic properties of
the predicted binary compounds.

\subsection{Electronic Properties}
\begin{figure*}
	\centering
	\includegraphics[width=0.9\textwidth]{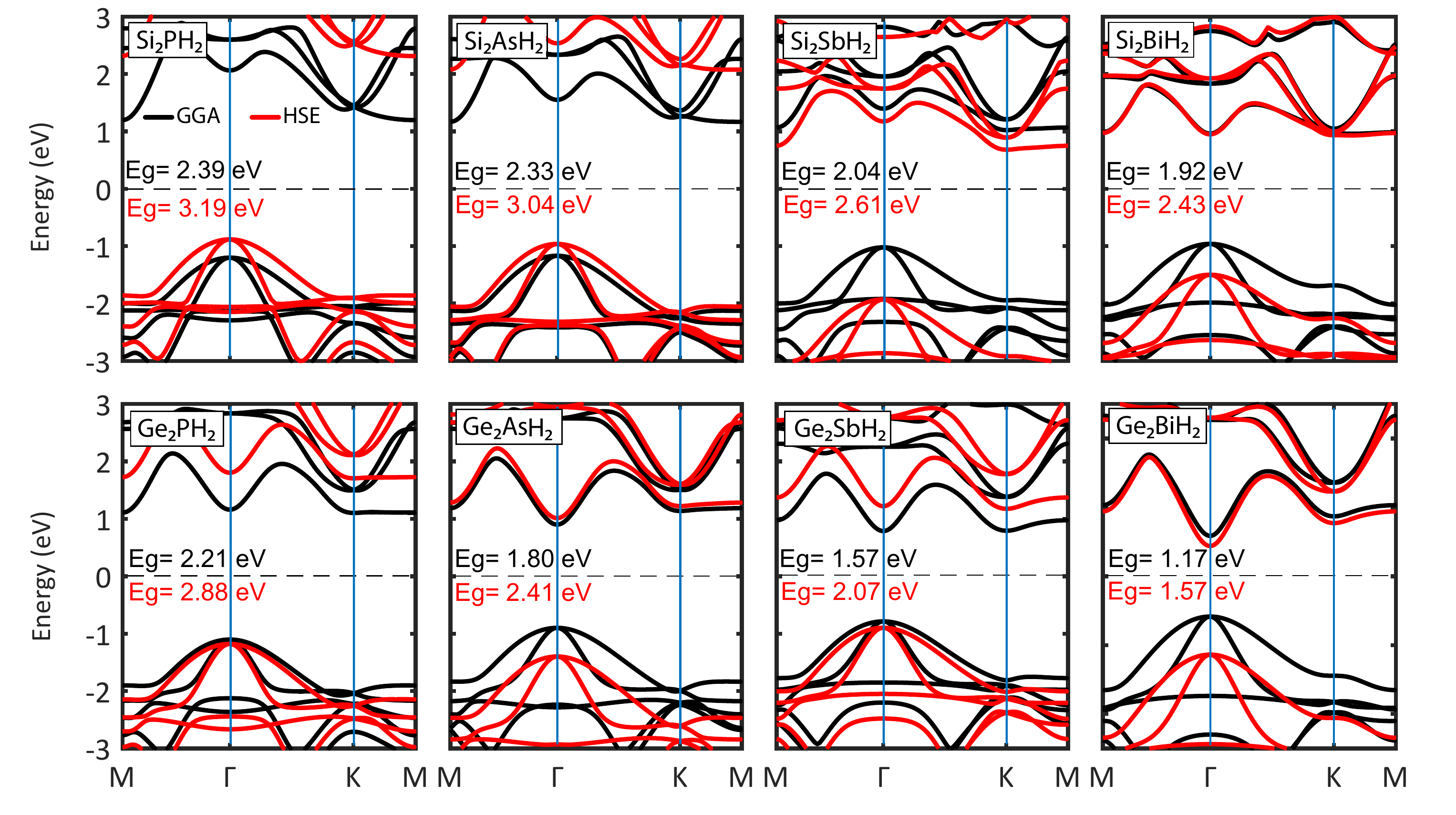}
	\caption{(Color online) Energy band structures of the
		X$_2$YH$_2$ binary compound monolayers along the main high symmetry
		k-points
		at
		the GGA (black lines) and HSE06 (red lines) levels together with the
		band
		gap values. The Fermi levels were shifted to zero.}
	\label{band}
\end{figure*}

The electronic band structures of X$_2$YH$_2$ binary compound monolayers have 
been
presented at the GGA and HSE06 levels in Fig.~\ref{band}. As can be seen, 
all
the
monolayers are semiconductors. The Ge$_2$YH$_2$ monolayers have direct band 
gaps 
at
the $\Gamma$ point. In contrast, the Si$_2$YH$_2$ monolayers have indirect band 
gaps
where their valence band maxima (VBM) are located at the $\Gamma$ point and
their conduction band minima (CBM) are located at the M (for Si$_2$PH$_2$ and
Si$_2$AsH$_2$) and K (for Si$_2$SbH$_2$ and Si$_2$BiH$_2$) points, which are 
identical at 
both
GGA and HSE levels. The band gaps predicted at the HSE level are in the 
range
of 1.57 to 3.19 eV, where Si$_2$PH$_2$ and Ge$_2$BiH$_2$ exhibit the largest and
smallest
values, respectively (see Table~\ref{configtab}). It is obvious that the 
band
gaps
decrease
regularly with increasing the average atomic mass, which is rather common 
in 2D
semiconductors \cite{ozdamar18, Yu16}. For example, in group V binary 
compound
monolayers, studied by Zhang et al, the
PAs and SbBi monolayers indicate the largest (2.55 eV) and smallest (1.41 
eV)
band gaps,
respectively. In more details, the reported band gaps are in the order of
PAs $>$ PSb $>$ PBi $>$ AsBi $>$ SbBi \cite{Zhang182}.

All the calculated band structures demonstrate parabolic valence bands 
centered at the $\Gamma$ point which provides high hole conductivity. Among 
these, 
the
Ge$_2$YH$_2$ structures have parabolic conduction bands centered at $\Gamma$ 
point,
which indicate free electrons, while the Si$_2$YH$_2$ structures have nearly 
flat
conduction bands along the K $-$ M direction, which are signatures of
localized
electrons. In other words, the Si$_2$YH$_2$ structures have both free and 
strongly
localized charge carriers like the Sn$_2$Bi monolayer deposited on the 
silicon
substrate \cite{gou18}. 

This high electron-hole asymmetry enforces the
materials to exhibit completely different optical and thermoelectric behavior 
in the n-type and p-type doping. In addition, all the monolayers have some 
conduction band
extrema (CBE) near the CBM at high symmetry points M, $\Gamma$, and K which 
may be
favorable for an n-type Seebeck coefficient \cite{Zhang17}. These CBEs may
approach each other by mechanical strain to achieve band convergence 
\cite{guo16}. The 
band convergence improves electrical conductivity without affecting other
transport coefficients. These features would make the X$_2$YH$_2$ monolayers 
possible candidates for thermoelectric applications.

\begin{figure*}
	\centering
	\includegraphics[width=0.9\textwidth]{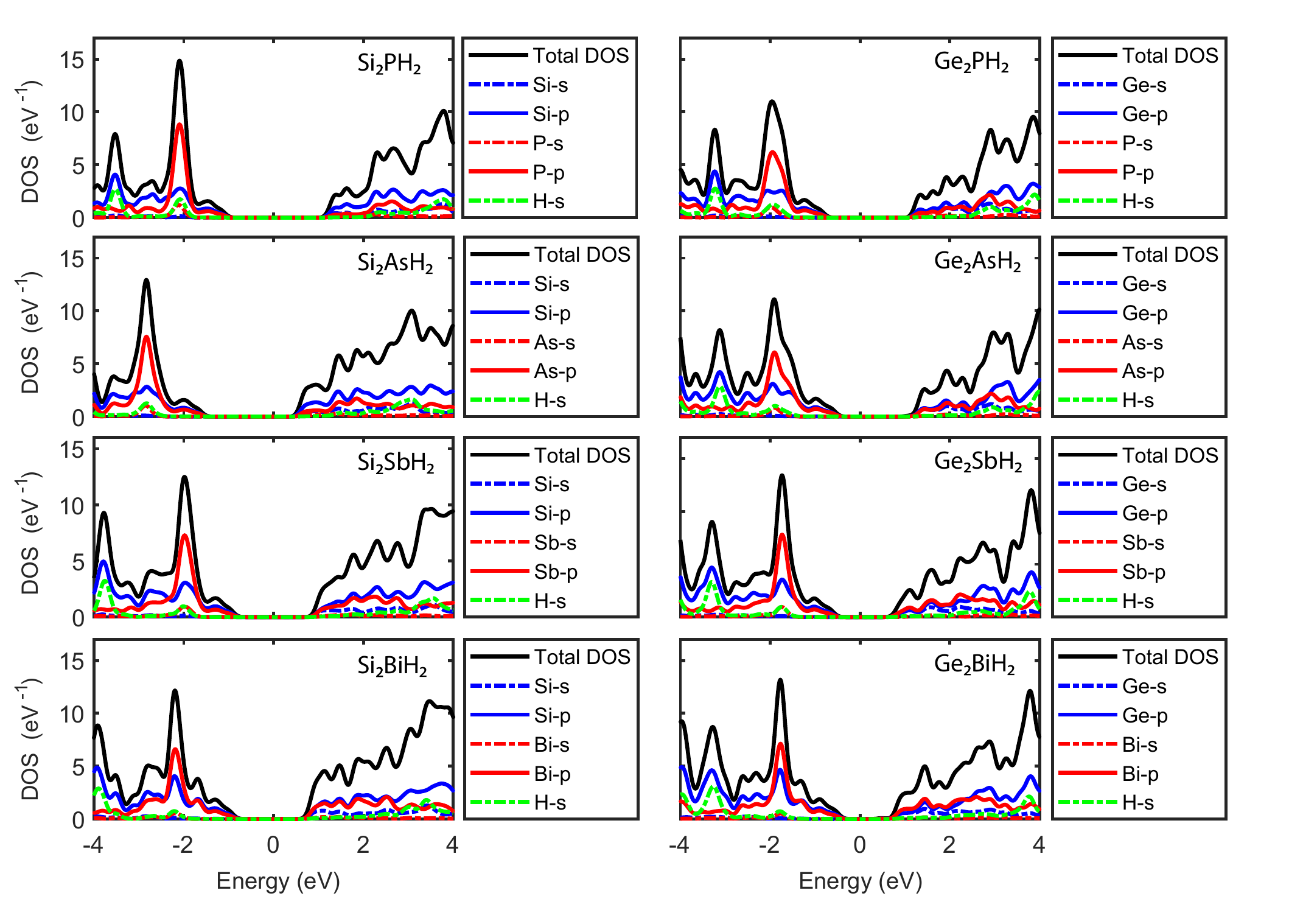}
	\caption{(Color online) Total and partial density of states of the
		X$_2$YH$_2$ binary compound monolayers at the GGA level. 
		The Fermi
		levels were shifted to zero.}
	\label{dos}
\end{figure*}

\begin{figure*}
	\centering
	\includegraphics[width=0.95\textwidth]{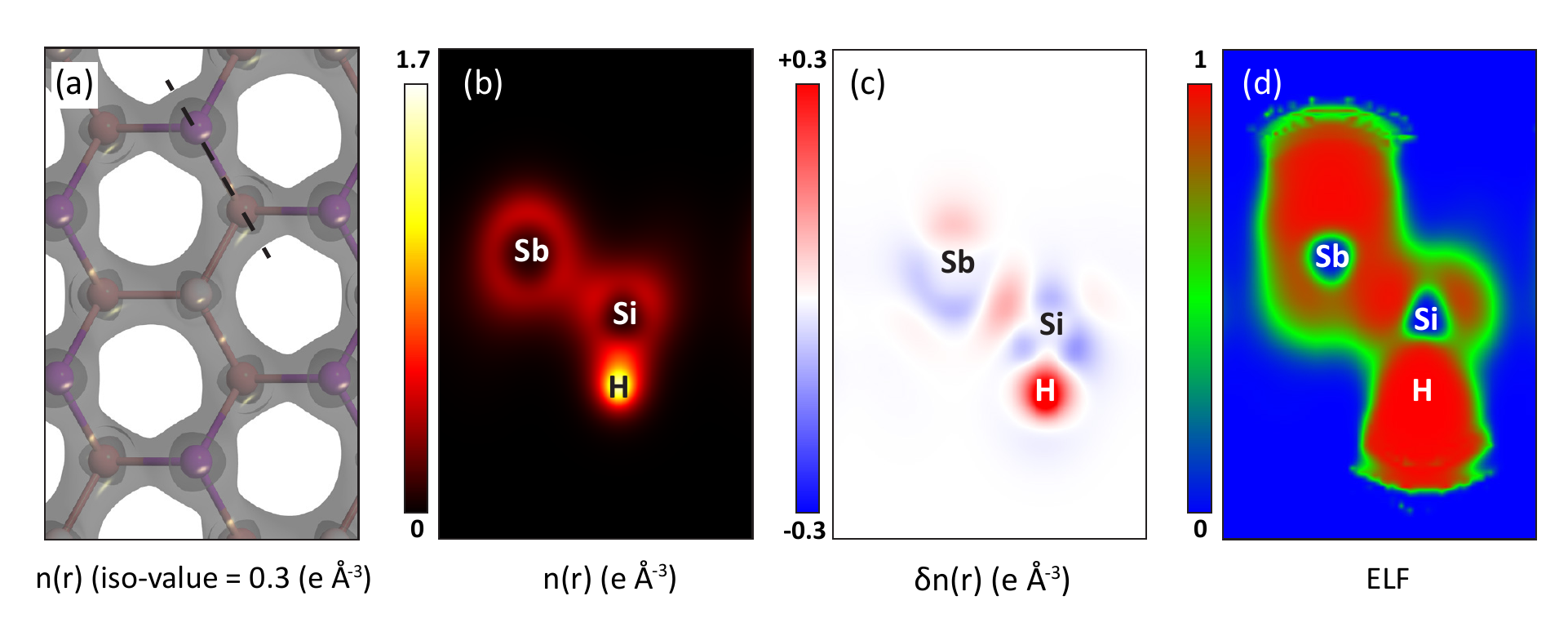}
	\caption{(Color online) Iso-surface and cut \textcolor{black}{plane} of
		the Si$_2$SbH$_2$ monolayer: \textbf{(a)} Iso-surface of
		electron density
		with
		an iso-value of 0.3
		$e$\AA$^{-3}$. \textbf{(b)} Cross-section cut 
		\textcolor{black}{plane} 
		of electron 
		density,
		\textbf{(c)}
		electron
		difference density, and
		\textbf{(d)} electron localization function (ELF)
		along the black dashed line in (a). The color bar next
		to the
		sub-figures
		denotes the scope of each quantity. In (c), the red and blue colors 
		show
		electron accumulation and depletion, respectively.}
	\label{elf}
\end{figure*}

We also took into account the spin-orbit coupling (SOC) interaction in the
calculation of the GGA band structures (SOGGA) as presented in Fig. S2. It
can be seen that consideration of the SOC, more or less, terminates the
degeneracy between energy bands and narrows the band gaps. Due to the 
stronger
spin-orbit interactions for heavier atoms, the band splitting increases as 
the
compounds are heavier.

Summarily, the effect of SOC on the band gaps is 
smaller
than 0.3 eV for most of the monolayers, except for relatively heavy X$_2$BiH$_2$
(Si$_2$BiH$_2$ \& Ge$_2$BiH$_2$) which have SOGGA band gaps approximately 0.5 
eV 
smaller
than that of GGA. Overall, for its small influence on most of the 
monolayers,
the SOC was not considered for the rest of our calculations.
Fig.~\ref{dos} shows the total and orbital projected density of states of 
the
hydrogenated binary monolayers. It can be seen that in the whole energy 
range,
the
p orbitals are dominant and the s orbitals have negligible proportions in 
the
electronic characteristics, which was predictable according to the 
electronic
arrangement of the contained atoms. This domination have been reported for
other group IV and V 2D structures \cite{Wang15,
	kamal15,akturk15, cahangirov09}. As it is clear, for all the
monolayers, the Y-p orbitals are dominant in the valance bands, and major 
peaks
around -2 eV are raised by them. These are attributed to the rather flat 
bands
around -2 eV in the band structures (see Fig.~\ref{band}). On the other 
hand,
the
conduction bands are slightly dominated by Si atom for the Si contained
structures, while for the Ge contained ones, Ge-p and Y-p orbitals share
rather equal proportions of the conduction band states.

Also, more or less, we see an overlapping of DOS of X-p and Y-p orbitals 
near
the Fermi energy for all the monolayers, which are signatures of strong
covalent bonds between the atoms, due to the orbitals hybridization. As can 
be
seen, orbitals hybridization is rather similar for all the compounds in the
valance bands, but in the conduction bands, it is not significant for
Si$_2$PH$_2$
and Si$_2$AsH$_2$. More interestingly, in the Ge contained compounds, the Y-s
orbitals
also participate in the hybridization. Orbitals hybridization between 
different
atoms was also reported for other structures such as Sn$_2$Bi, C$_3$N, 
C$_3$P,
and C$_3$As compounds \cite{ding19, li19-RYwTi}.

Moreover, the H atoms have a very limited contribution in the DOS, which 
means
that electrons are strongly bound to them and do not construct many states
in the valance and conduction bands. Namely, a very small hybridization with
Y-s orbitals, and no interfere with \mbox{X orbitals} is seen, which
suggests
ionic
bonds between the H and X atoms.

To shed more light on the electronic properties and bonding mechanism of
the compounds, electron density ($n(r)$), electron difference density
($\delta n(r)$), and electron localization function (ELF) were calculated at
the
GGA level. Our calculations display that all the monolayers have similar
characteristics, therefore, we only present the analyses for Si$_2$SbH$_2$
monolayer,
as a representative, in Fig.~\ref{elf}. Analyses for the rest of the
monolayers
are available in Fig. \mbox{S3 $-$ S5.} It is clear from Fig. 
\ref{elf}a,
that
the
lattice
has
a minimum uniform electron density of about 0.3 e \AA$^{-3}$ which exhibits 
an
in-plane isotropic lattice in aspects of electronic characteristics. It is
obvious from $n(r)$ and $\delta n(r)$ (Fig.~\ref{elf}b,c) that there is a
gentle
electron accumulation between Sb and Si atoms. Moreover, the ELF (Fig.
\ref{elf}d)
indicates a high localization between these atoms. Therefore, one could
conclude that the Sb and Si atoms share electrons mutually and make covalent
bonds.

Meanwhile, there is a high electron density and accumulation on the H,
with significant electron depletion around Si atoms. Besides, the ELF 
displays
the highest localization on the H and a low localization around the Si 
atoms.
Therefore, it is deducible that the H atoms make ionic bonds with Si atoms.
This approves our discussion about the low contribution of H related 
electrons
in the density of states. Also, the strong ionic bonds make sense about the
stability of the monolayers after hydrogenation. In other words, the
hydrogenation somehow plays the role of a substrate for the originally 
unstable
pristine monolayers and stabilizes them.

\subsection{Optical Properties}

\begin{figure*}
	\centering
	\includegraphics[width=0.85\textwidth]{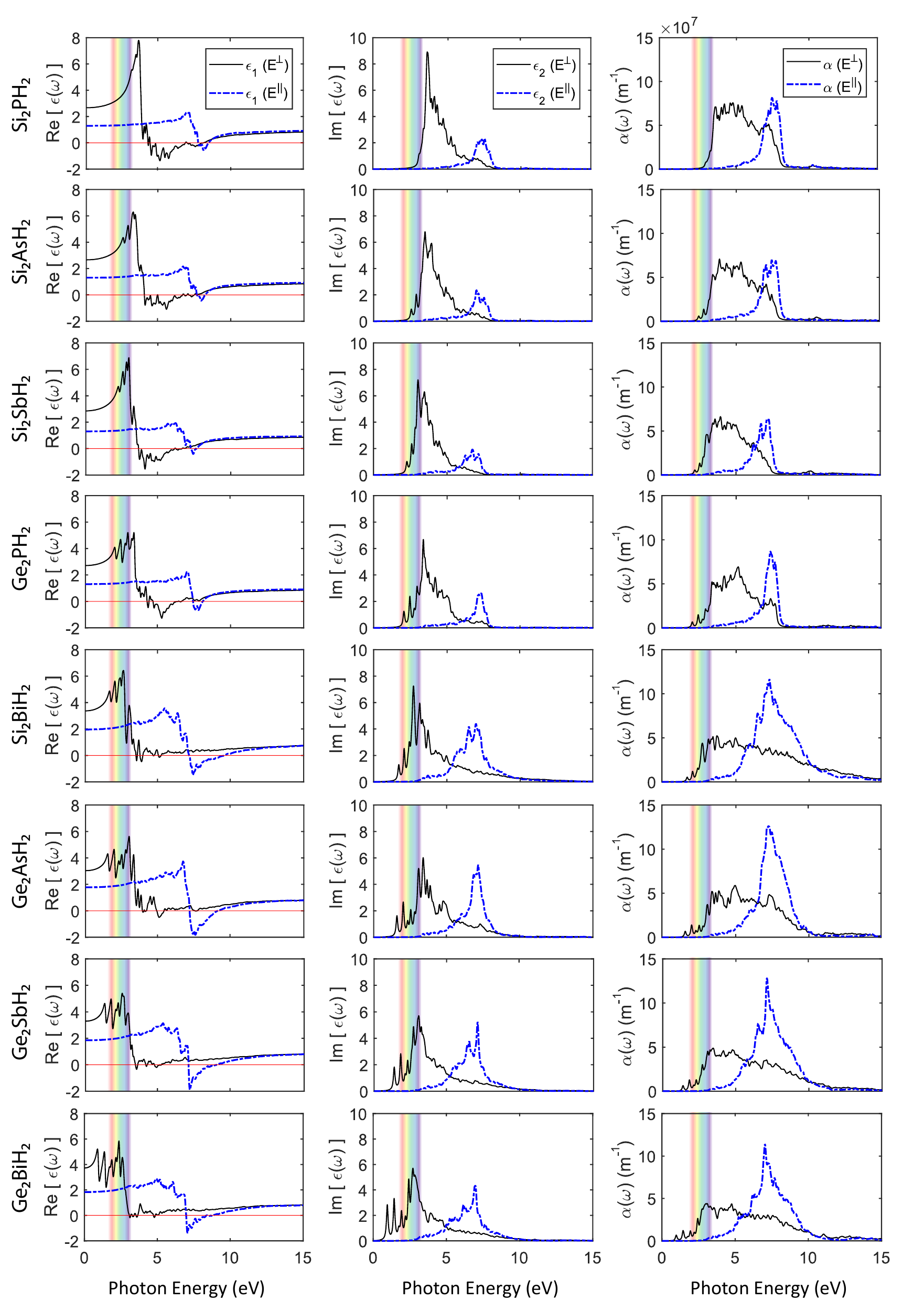}
	\caption{Optical properties of the X$_2$YH$_2$ binary compound
		monolayers, including the real and imaginary parts of the dielectric
		function ($\epsilon_1$ and $\epsilon_2$), and absorption coefficient
		($\alpha$), for the perpendicular (E$^\perp$) and parallel
		(E$^\parallel$)
		polarizations.}
	\label{optic}
\end{figure*}

\begin{table*}
	\footnotesize
	\centering
	\caption{Optical parameters of the X$_2$YH$_2$ binary compounds,
		for
		perpendicular and parallel polarizations.
		\mbox{Metallic Range} Stands for the range in which the real
		part of
		the dielectric function has negative values, and the monolayer is
		metallic.
		E ($\epsilon_2^{peak}$) represents the energy in which the imaginary
		part of the dielectric function has the major peak. Sig. $\alpha$
		Rng. shows the range in which the optical absorption coefficient has
		significant values (e.g. $\alpha \geq 10^{7} m^{-1}$). The mentioned
		ranges
		are rounded
		to the nearest 5 nm for more clarity.}
	\label{optictab}
	
	\begin{tabular}{ccccc|ccc}
		\hline
		&  & \multicolumn{3}{c|}{E$^\perp$} &
		\multicolumn{3}{c}{E$^\parallel$}  \\
		\cline{3-5}\cline{6-8}
		& & Metallic Range (nm) & $E (\epsilon_2^{peak})$ (eV) &
		High Abs. Range (nm) & Metallic Range (nm) & $E 
		(\epsilon_2^{peak})$
		(eV) &
		Sig.
		$\alpha$
		Rng. (nm)                   \\
		\cline{3-5}\cline{6-8}
		\multirow{4}{*}{\rotatebox{90}{Group-A\;\;}}
		&Si$_2$PH$_2$	&280 $-$ 155	&3.74 	&375 $-$ 150		& ---	
		&7.69	
		&195 $-$ 145                 \\
		&Si$_2$AsH$_2$	&310 $-$ 165	&3.53	&395 $-$ 155		& ---	
		&7.06	
		&200 $-$ 150                 \\
		&Si$_2$SbH$_2$	&340 $-$ 200	&3.05	&465 $-$ 170	& ---	&6.77	
		&220 $-$ 160\\
		&Ge$_2$PH$_2$	&290 $-$ 195 	&3.4	&430 $-$ 155	& ---	&7.37	
		&210 
		$-$ 155\\
		\cline{2-5}\cline{6-8}
		\multirow{4}{*}{\rotatebox{90}{Group-B\;\;}}
		&Si$_2$BiH$_2$	& ---	&2.72	&525 $-$ 100	&175 $-$ 130	&7	
		&250 $-$ 125	
		\\
		&Ge$_2$AsH$_2$	& ---	&3.38	&450 $-$ 135	&170 $-$ 140	&7.15	
		&250 $-$
		125	\\
		&Ge$_2$SbH$_2$	& ---	&3.05	&495 $-$ 120	&175 $-$ 135	&7.07	
		&255 $-$
		120	\\
		&Ge$_2$BiH$_2$	& ---	&2.72	&530 $-$ 120	&180 $-$ 140	&6.98	
		&270 $-$
		125	\\
		\hline
	\end{tabular}
	
\end{table*}

High optical absorption in 2D materials brings hopes for energy
harvesting purposes such as solar cells. Moreover, linear dichroism is a
phenomenon widely reported for 2D materials, which is the
difference between optical absorption for light
beams polarized parallel and perpendicular to an orientation axis, and is a
key element for interesting optical applications such as beam splitters, 
LCDs,
half-mirrors, etc. \cite{Wang19, xu17}. For instance, it 
is
reported that Sb and As monolayers have
optical absorption edges near $\sim$ 2 and $\sim$ 3 eV , for perpendicular
and parallel
polarizations, respectively \cite{xu17}.

As mentioned in the previous section,
the X$_2$YH$_2$ binary compounds were predicted to have hopeful
signs of optical potentials, such as wide band gaps in the range of visible
light. In this section, we calculate and discuss the optical properties of
these
monolayers to extract more physical insights and possible applications.

The optical properties are associated with the interactions between light,
electrons, and ions in the materials, which should be explained through the
complex dielectric function,
$\epsilon(\omega)=\epsilon_1(\omega)+i\epsilon_2(\omega)$. Based on Fermi’s
golden rule, one can derive the imaginary part of the
dielectric function as below \cite{gajdos06}: \vspace{-2pt} 

\begin{equation}
\epsilon_2(\omega)=\frac{4\pi^2e^2}{m^2\omega^2}\sum_{C,V}\left| P_{C,V}
\right|^2\delta \left( E_C - E_V - \hslash\omega\right)
\end{equation}

\noindent where $e$ is the electron charge, $m$ is the electron effective 
mass,
$P$ is
the
momentum transition matrix, and $E$ is the electron energy level. Moreover, 
$C$
and
$V$
indices stand for conduction and valance bands, respectively. No need to
explain, $\delta(x-x_0)$ is the Dirac delta function, which ensures
conversion of energy during electron transitions from band to band. This 
means that every excited state has an infinite lifetime, i.e. is 
stationary \cite{benassi08}. Subsequently, the
real part can be calculated through Kramer-Kronig relation: 
\cite{carcione19}

\begin{equation}
\epsilon_1(\omega)=1+\frac{2}{\pi}\int_{0}^{\infty}\frac{\omega'\epsilon_2(\omega')}
{\omega'^2-\omega^2}d\omega
\end{equation}

Moreover, based on the real and imaginary parts, the optical absorption
coefficient, $\alpha(\omega)$, is calculated through:

\begin{equation}
\alpha(\omega)=2\frac{\omega}{c}\sqrt{\frac{\sqrt{\epsilon_1^2+\epsilon_1^2}-\epsilon_1}
	{2}}
\end{equation}

\noindent where $c$ is the speed of light.

For the isotropy of the monolayers in the xy plane, there is no significant
difference between xx and yy polarizations, therefore the calculations were
performed for polarized radiations, parallel (E$^\parallel$) and 
perpendicular
(E$^\perp$) to
the incidence direction (z-direction). \mbox{Fig.~\ref{optic}} shows the
calculated
optical
properties of the X$_2$YH$_2$ binary compounds, including real and
imaginary parts of the dielectric function ($\epsilon_1$ and $\epsilon_2$), 
and
the absorption coefficient ($\alpha$), for both polarizations. 
Interestingly,
the predicted monolayers can be separated into two groups, group-A, 
including
Si$_2$PH$_2$, Si$_2$AsH$_2$, Si$_2$SbH$_2$, and Ge$_2$PH$_2$, and group-B 
including 
Si$_2$BiH$_2$,
Ge$_2$AsH$_2$, Ge$_2$SbH$_2$, and
Ge$_2$BiH$_2$. The materials in each group exhibit similar properties, which 
will be
discussed in detail.

As we know, negative values in the real part of the dielectric function
stand for metallic reflectivity \cite{singh16}. As it is clear in Fig.
\ref{optic}
(left panel), group-A monolayers have significant negative values in the
real part of the dielectric function within $\sim$ 3.6 to 8 eV ($\sim$ 345 
to
155 nm) in the UV region, for perpendicular polarized radiation 
(E$^\perp$). On
the contrary, group-B monolayers have significant negative values within
$\sim$ 6.8 to 9.5 eV
($\sim$ 180 to 130 nm), for parallel polarized radiation (E$^\parallel$). In
other words, group-A and group-B materials are metallic for E$^\perp$ and
E$^\parallel$ UV radiation, within the mentioned ranges, respectively. This
means that group-A and group-B monolayers have a good complement in blocking
the
UV radiation and may be used together as a heterostructure for more 
efficient
beam splitting, and UV protection purposes. Compared with the
Si and Ge monolayers, which have been
reported to have a metallic characteristics in the range of $\sim$ 4 to 7 eV
(310 to 177 nm)
and $\sim$ 0 to 4 eV ($\infty$ to 310 nm), respectively \cite{chen16, 
	mohan14},
most of the predicted X$_2$YH$_2$ binary compounds have better UV blocking. For 
more details, please see Table~\ref{optictab}.

The imaginary part of the dielectric function and the absorption coefficient
are bound to each other and should be analyzed together. Based on the band 
to
band transition theory, the peaks in the imaginary part of the dielectric
function are concerned with energy absorption and direct transitions of
electrons between bands below and above the Fermi level. As can be seen in
Fig.~\ref{optic} (middle panel), all the monolayers have major peaks 
around
$\sim$~3.5
and $\sim$ 7 eV for E$^\perp$ and E$^\parallel$ polarizations, respectively.

Moreover, in group-A monolayers, the E$^\perp$ peaks are much stronger than 
the
E$^\parallel$ peaks, whereas, in group-B monolayers, they are relatively 
equal.
This would be representative of the difference, and equality of significant
absorption ranges \mbox{($\alpha \geq 10^{7} m^{-1}$)} between E$^\perp$ and
E$^\parallel$
polarizations, for
group-A and group-B monolayers, respectively. In other words, as it is 
shown in
Fig.~\ref{optic} (right panel), group-B monolayers have relatively wider
significant absorption ranges for E$^\parallel$ polarizations, which is due 
to
the stronger E$^\parallel$ peaks in the imaginary part of the dielectric
function.

The widest significant absorption range belongs to Si$_2$BiH$_2$, which is in 
the
range of 2.36 to 12.4 eV (525 to 100 nm). For comparison, it should be noted
that
the Si and Ge monolayers have significant optical absorption in the range of
$\sim$ 3.5 to 5 eV (354 to 248 nm) and $\sim$ 3 to 6 eV (413~to~206.6~nm),
respectively \cite{chen16, mohan14}. Our calculations show that most of the
predicted compounds have greatly wider significant absorption ranges. For 
more
details
about the optical properties, please note to \mbox{Table~\ref{optictab}}.

Summarily, one can conclude that group-A monolayers, having stronger linear
dichroism, have more potential applications in beam splitting, and group-B
monolayers, having a wider absorption range for both polarizations, are more
favorable for energy harvesting systems and solar cells. It should be added
that three of group-B monolayers, namely Ge$_2$AsH$_2$, Ge$_2$SbH$_2$, and 
Ge$_2$BiH$_2$
have direct and wide band gaps, which makes them even more ideal for this
purpose.

\subsection{Photocatalytic Properties}

\begin{figure}
	\centering
	\captionsetup{justification=centering}
	\includegraphics[width=0.5\textwidth]{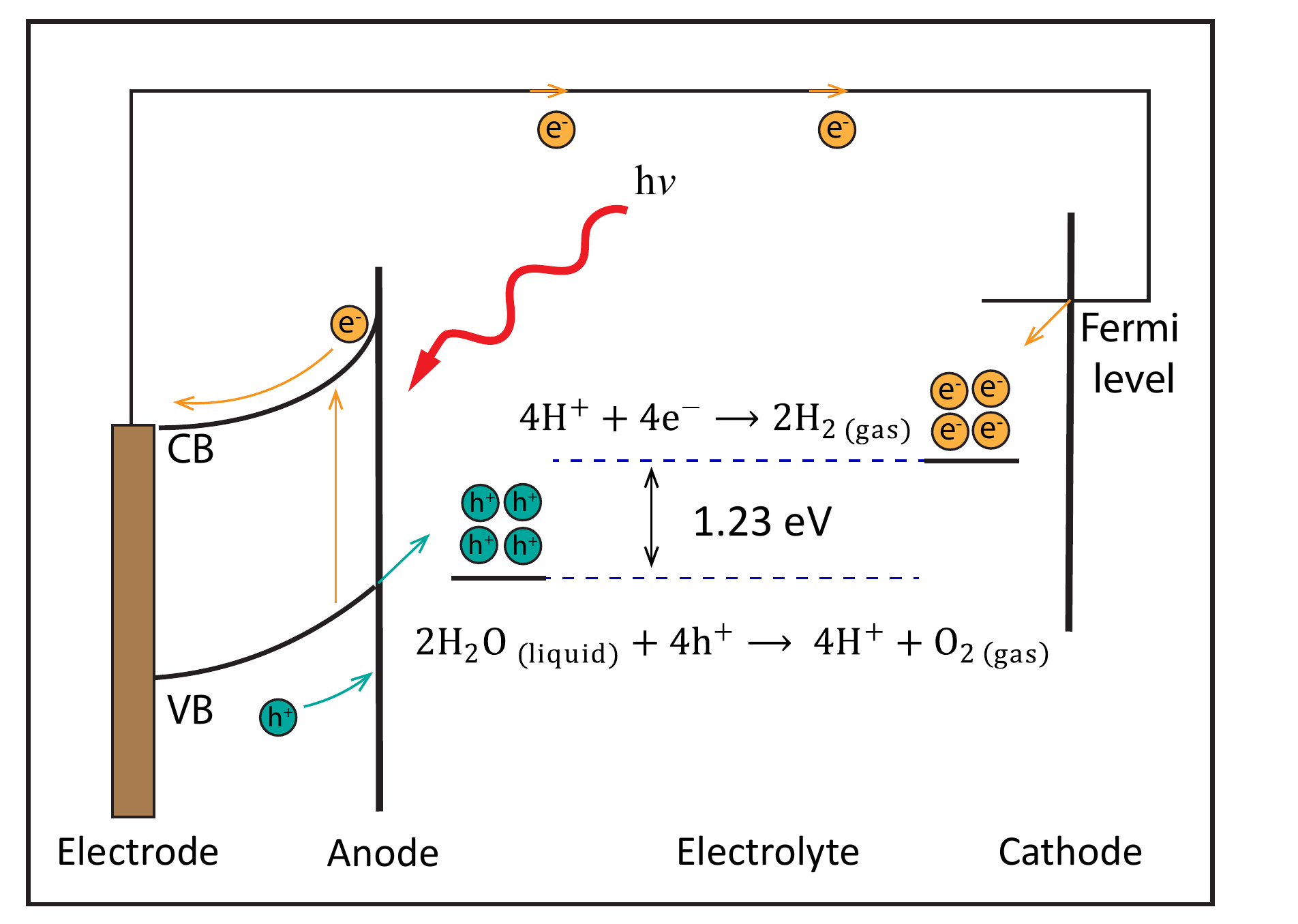}
	\caption{A schematic of photocatalytic water splitting process.}
	\label{shceme}
\end{figure}

Water splitting is a chemical reaction in which the water molecule is broken
down into oxygen and hydrogen. This process has attracted much
attention
because of clean, inexpensive, and environment friendly production of
hydrogen. One of the well-known methods for water 
splitting is photocatalysis by use of a semiconductor sheet and solar energy 
\cite{chen2016recent}.  
The general chemical formula for this reaction is presented as \cite{jeong18}:

The first half reaction shows the water oxidation at the anode and the 
second
one indicates the water reduction at the cathode. The overall process 
results
in
production of hydrogen and oxygen gases as illustrated in Fig.
\ref{shceme}. A semiconductor could be a potential photocatalyst for water
splitting if the
CBM energy is higher than the reduction potential of H$^+$/H$_2$, and the 
VBM
energy is lower than the oxidation potential of O$_2$/H$_2$O 
\cite{Zhu19_a}. It
should be noted that there are not many photocatalysts that meet all of the
requirements, so far. Therefore, finding a suitable candidate semiconductor 
for
this purpose is a crucial challenge, that we are going to face in this
section.
\hspace{-2pt}
\begin{equation}
\begin{gathered}
\mathrm{2H_2O_{(liquid)}+4h^+\rightarrow 4H^+ + O_{2 (gas)}}\\
\mathrm{4H^+ +e^- \rightarrow 2H_{2 (gas)}}
\end{gathered}
\end{equation}

\begin{figure}
	\centering
	\includegraphics[width=0.5\textwidth]{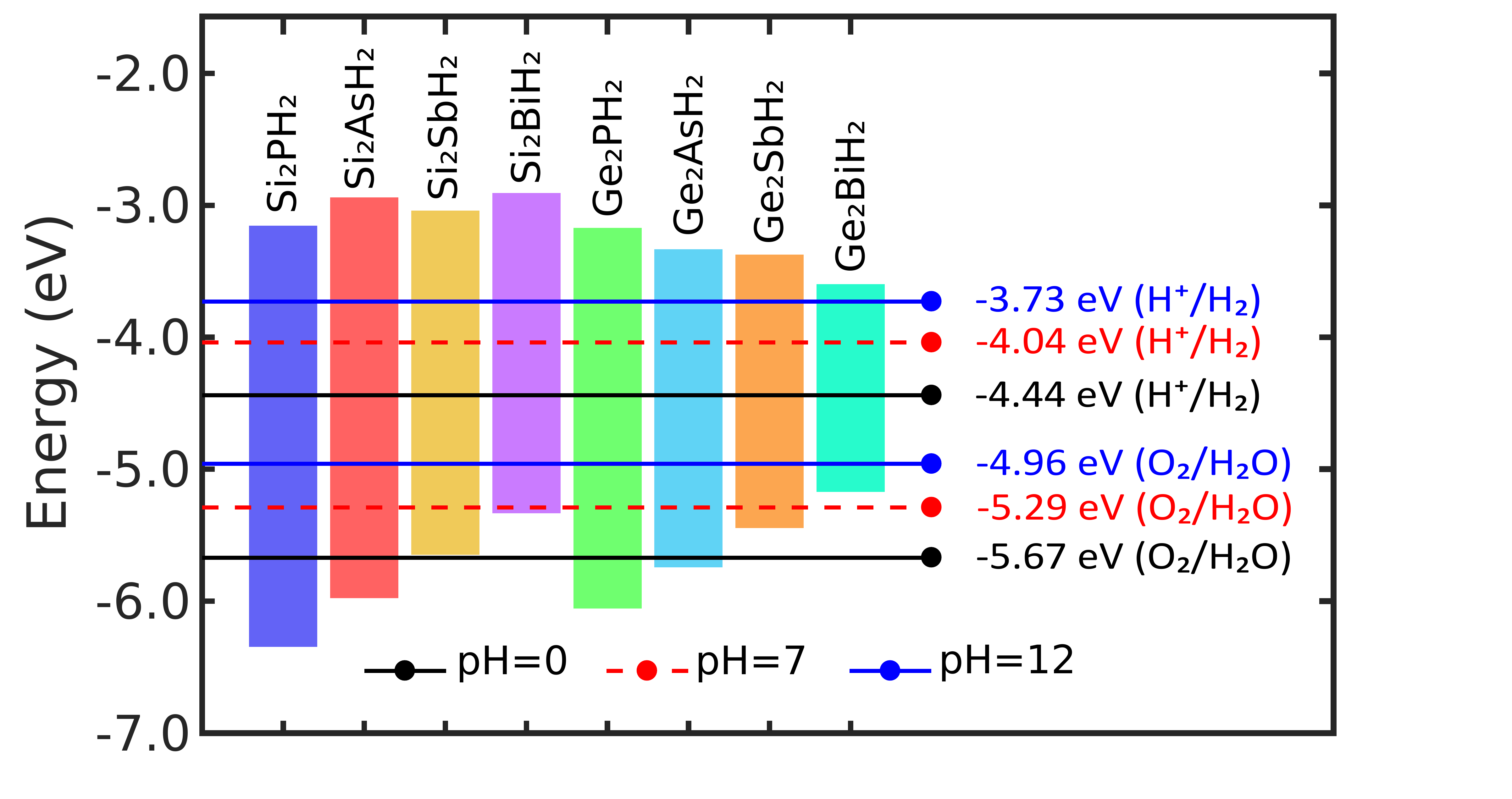}
	\caption{(Color online) Band edge positions of X$_2$YH$_2$ binary compound
		monolayers for
		photocatalytic water splitting, calculated at the HSE06 level. The 
		redox
		potentials of water splitting reaction have been specified at the 
		pH=0
		(black solid lines), pH=7 (red dashed lines), and pH=12 (blue solid 
		lines).}
	\label{water}
\end{figure}

Due to dependency of the reduction/oxidation (known as redox) potentials to 
the pH, these potentials were adopted at pH=0, 7, and 12, similar to the
previous studies \cite{chowdhury17, Zhang18-0}. In Fig.~\ref{water}, the HSE06 
band
edges of the
X$_2$YH$_2$
monolayers have been presented with respect to the vacuum level. As can be
seen, at pH=0, the X$_2$PH$_2$ and X$_2$AsH$_2$ monolayers have suitable band 
edge 
for
water splitting reaction while at pH=12, all the monolayers are eligible.
However, this reaction usually occurs in a neutral environment (pH=7). At
this
pH, all the monolayers except Ge$_2$BiH$_2$ satisfy the condition of the band 
edge
position.

As suggested by Zhang et al, materials with indirect band gaps are more
desirable for photocatalytic activity \cite{Zhang14}, therefore
Si$_2$YH$_2$
monolayers will
react better than Ge$_2$YH$_2$ ones. On the other hand, the band gap value 
should be
smaller than 3 eV for enhancing the visible light absorption 
\cite{Yang18,liao13},
therefore the Si$_2$PH$_2$ and Si$_2$AsH$_2$ monolayers, having large band gaps 
for
visible light, cannot produce high efficiency for electron-hole generation
and accordingly for water splitting. Summarily, the Si$_2$SbH$_2$ and 
Si$_2$BiH$_2$
monolayers are very promising candidates for water splitting.

\section{Conclusion}
In summary, using first-principles calculations, for the first time, we have
proposed a new family
of two-dimensional binary compounds with an empirical formula of X$_2$Y, 
where
X and Y belong to groups IV (Si and Ge) and V (P, As, Sb, and Bi),
respectively. Different from their pure structures, the hydrogenated 
(X$_2$YH$_2$) monolayers exhibit a very high stability according to cohesive 
energy,
phonon dispersion analysis, and AIMD simulations. We have obtained many
interesting physical properties by computing the electrical, optical, and 
photocatalytic behavior of these monolayers. Our calculations
disclose that all of the monolayers are semiconductors with band gaps in
the range of 1.57 to 3.19 eV. The optical results reveal that Si$_2$PH$_2$,
Si$_2$AsH$_2$, Si$_2$SbH$_2$, and Ge$_2$PH$_2$
monolayers
have potential applications in beam splitting, and Si$_2$BiH$_2$, Ge$_2$AsH$_2$,
Ge$_2$SbH$_2$, and Ge$_2$BiH$_2$ monolayers are
more favorable for energy harvesting systems and solar cells. Besides, the 
Si$_2$SbH$_2$ and Si$_2$BiH$_2$ monolayers were found to have suitable 
band
gaps and band edge positions for photocatalytic water splitting. Our results
suggest the binary monolayers of group IV-V for uses in nano-electronic and
optoelectronic applications, and propose them for further
experimental works. \textcolor{black}{Finally, we predict that the reported 
	properties 
	for X$_2$YH$_2$ monolayers would be also similar to possible deposited 
	X$_2$Y monolayers  on a proper substrate such as Si (111), ZnS (111), and 
	SiC 
	(111).}

\section*{Supporting Information}

See the Supporting Information for details about the structural and 
electronic properties.

\section*{Conflicts of Interest}

The authors declare that they have no conflict of interest.

\section*{Acknowledgment}

We are thankful to the Research Council of University of Guilan for the
partial support of this research.


\bibliographystyle{naturmag}
\bibliography{Abbr} 

\end{document}